\documentclass[conference]{IEEEtran}
\IEEEoverridecommandlockouts
\usepackage[bookmarks=false]{hyperref}
\usepackage{cite}
\usepackage{amsmath,amssymb,amsfonts}
\usepackage{algorithmic}
\usepackage{graphicx}
\usepackage{textcomp}
\usepackage{xcolor}
\usepackage{enumitem}
\usepackage[ruled, vlined, algo2e]{algorithm2e}

\usepackage{multirow}
\usepackage{array}
\usepackage{makecell}
\usepackage{booktabs}
\usepackage{placeins}

\usepackage{comment}

\usepackage{stfloats}

\usepackage{fancyhdr}
\fancypagestyle{preprintbanner}{%
  \fancyhf{}%
  \fancyhead[C]{\footnotesize Accepted for publication in the IEEE Journal of RFID. This is the preprint version of the accepted manuscript.}%
}
\usepackage{algorithm}
\usepackage{algorithmic}

\renewcommand{\arraystretch}{1.5}

\def\BibTeX{{\rm B\kern-.05em{\sc i\kern-.025em b}\kern-.08em
    T\kern-.1667em\lower.7ex\hbox{E}\kern-.125emX}}
\begin{document}

\makeatletter
\newcommand{\linebreakand}{%
  \end{@IEEEauthorhalign}
  \hfill\mbox{}\par
  \mbox{}\hfill\begin{@IEEEauthorhalign}
}
\makeatother

\title{Is Forecasting Accuracy Enough? A Comparative Study of Traffic Forecasters for Beam-Hopping LEO Satellite Networks\\
}

\author{\IEEEauthorblockN{Yekta Demirci\IEEEauthorrefmark{1},
% Olfa Ben Yahia\IEEEauthorrefmark{1},
Guillaume Mantelet\IEEEauthorrefmark{2},
Stéphane Martel\IEEEauthorrefmark{2}}
\IEEEauthorblockN{
Jean-François Frigon\IEEEauthorrefmark{1},
Gunes Karabulut Kurt\IEEEauthorrefmark{1}}
\IEEEauthorblockA{\IEEEauthorrefmark{1}Poly-Grames Research Center, Department of Electrical Engineering, Polytechnique Montréal, QC, Canada}
\IEEEauthorblockA{\IEEEauthorrefmark{2} Satellite Systems, MDA Space, Canada}
}

\maketitle
\thispagestyle{preprintbanner}

\begin{abstract}
We evaluate diverse models for user traffic demand forecasting in Low Earth Orbit (LEO) satellite networks 
with Beam Hopping (BH), questioning whether predictive accuracy is the right objective for this task. 
To capture the complex nature of the user traffic demand, we employ a second-order self-similar traffic model, 
supplemented by a publicly available Wi-Fi dataset to validate the self-similar model against the empirical 
traffic patterns. We compare forecasters ranging from classical statistical approaches, such as the optimal 
forecaster for self-similar data and the optimal linear predictor on the discrete sampling grid, to Fractional 
Auto-Regressive Integrated Moving Average (FARIMA) models, as well as emerging deep learning architectures. 
The latter category encompasses foundation and domain-specific transformer models, alongside a lightweight 
neural network consisting solely of linear layers. We assess these models at two levels: in isolation, 
through the Mean Absolute Scaled Error (MASE), and in context, through a BH simulator in which the forecast 
drives the illumination plan. On purely self-similar traffic the three self-similarity aware forecasters 
perform on par with one another and dominate the learned models, whereas on the raw Wi-Fi trace this 
ordering nearly reverses. Seasonality violates their stationary increment assumption; removing the periodic 
component restores their comparative accuracy. Crucially, these accuracy differences barely propagate to the 
system level. Loss ratio and buffer backlog are affected more by system utilization and the planning period than 
by the choice of forecaster, with the performance gap between forecasters vanishing entirely below 0.90 
utilization. This suggests design efforts are better spent optimizing utilization margins and planning periods 
rather than chasing marginal gains in raw accuracy.
\end{abstract}


\section{Introduction}
Low Earth Orbit (LEO) satellites are uniquely positioned to provide global connectivity through constellations formed by the dense meshes of interconnections. Even though the concept of LEO-based communications dates back to a few decades to pioneers like Globestar and Iridium \cite{prattConstel}, the industry has only recently entered a phase of rapid, large-scale deployments. The players such as Telesat, OneWeb, SpaceX (Starlink), and Amazon LEO (Kuiper) proposed deploying modern constellations consisting of thousands of satellites in iterative phases \cite{old}. These rapid deployments are also reflected by data where the annual launches have increased fivefold from 2019 to 2024, accompanied by an exponential rise in the cumulative number of satellites currently in orbit since 2019 \cite{fivefold}.

The deployments have been accelerated by a convergence of economic and technological factors. Primarily, the reusable rockets have fundamentally changed the economics of space access, reducing the launch costs more than threefold since the early 2000s \cite{adilovCost}, complemented by advancements in onboard processing and antenna technology, specifically the development of advanced reconfigurable phased arrays and electrically steerable beamforming systems \cite{yahia}. These innovations have provided the technical foundation for space-based solutions to help meet the escalating global demand for high-speed broadband \cite{yaacoub2020key} by supplementing terrestrial infrastructure.

Addressing the global demand via space-based solutions sparked a fundamental shift in satellite system design, moving from traditional bent-pipe architectures toward regenerative payloads. In a bent-pipe system, the satellite acts as a simple transparent relay, performing only RF filtering, amplification, and frequency conversion without accessing the underlying data. In contrast, regenerative payloads allow the satellite to demodulate and decode signals, enabling on-board routing and more granular signal management. Crucially, this regenerative capability supports multi-beam operations, allowing satellites to move beyond static footprints toward highly flexible resource allocation techniques, such as Beam Hopping (BH).

BH is a dynamic resource allocation technique that is used in multi-beam satellites to manage non-uniform user traffic demand. In a traditional multi-beam satellite, power and bandwidth are often distributed equally across different beams, regardless of actual demand. Yet the user demand is inherently nonuniform. For instance, a beam over a densely populated area is likely to require significantly more service capacity compared to open oceans with sparse aeronautical and maritime traffic. BH emerges as a promising tool to mitigate this problem by offering Time-Division Multiplexing (TDM) and adjustable Dwell-Time (DT) where the DT is illumination duration of a beam. Instead of illuminating all beams simultaneously, the satellite would ``hop" its beams between different areas in a predefined or adaptive time window. By dynamically adjusting the DT and the bandwidth for each beam, the system can proactively align satellite resources with the spatiotemporal fluctuations of the user traffic demand.

Yet this flexibility comes with a design cost. Despite the inherent flexibility of BH, its operational efficiency is strictly restrained by the design of the beam illumination plan. Therefore, a significant amount of research has focused on optimizing these plans from different perspectives. For instance, several studies have integrated BH scheduling with physical layer constraints, such as carrier aggregation \cite{kibria2022joint} and hybrid digital-analog beamforming to maximize system sum-rates \cite{wang2024joint}. Others have employed advanced mathematical and algorithmic frameworks, including mean-field theory for joint power and scheduling design \cite{chen2024joint} and reinforcement learning for autonomous plan generation \cite{ran2024towards}. Furthermore, the scope has expanded to multi-satellite coordination, addressing sub-problems like load balancing and inter-satellite interference avoidance \cite{lin2022multi}, as well as geometric optimizations that treat beam positioning as a p-center problem to minimize queue delays \cite{tang2021optimization}.

However, there remains a critical limitation in these approaches, the assumption of static or slowly-varying traffic demands. While the allocation methods focus on the optimization aspect, they simply use current traffic snapshots, yet in highly dynamic LEO environments, an ``optimal" solution may already be obsolete by the time it is implemented. While some frameworks acknowledge traffic variability \cite{xu2022joint}, they lack a mechanism to anticipate future demand. This creates a significant literature gap regarding predictive resource management. Specifically, there is a lack of research into how near-term traffic forecasting should be done, so that it can be integrated into the BH optimization loop to ensure that capacity allocation remains aligned with the actual spatiotemporal fluctuations of user demand.

Since the user demand is tracked by discrete temporal samples, this is essentially a Time Series Forecasting (TSF) problem, a field in which the research community has also proposed numerous solutions. Broadly the forecasters can be categorized into statistical and data-driven deep learning models. Traditional statistical models, such as the Autoregressive Integrated Moving Average (ARIMA) \cite{box1970}, rely on historical parameters to project future values and remain widely used in fields ranging from finance to cloud resource management \cite{10636792}. On the other hand, deep learning approaches leverage neural networks being pre-trained on high dimensional data to capture complex non-linear patterns. Some examples include Long Short-Term Memory (LSTM) networks \cite{8614252}, DeepAR \cite{SALINAS20201181} and Multi-Layer Perceptron based architectures like N-Beats \cite{nbeats} and DLinear \cite{Zeng2022AreTE}. More recently, Transformer based architectures have redefined State of the Art (SOTA) performance in TSF. Transformers are a neural network architecture consisting of an encoder and a decoder layers that utilize attention mechanism to model relationships between any two positions in a sequence, regardless of their distance \cite{vaswani2017attention}. Even though they were initially proposed for sequence to sequence problems, their strong pattern recognition capabilities soon applied for TSF. 

Transformer based models can be further divided into domain specific models such as Informer \cite{zhou2021informer} which is trained solely on a specific dataset and foundation models such as Chronos \cite{ansari2024chronos} and  TimesFM \cite{das2024decoder} which are pre-trained on massive, diverse datasets to enable zero-shot forecasting across any domain. While foundation models offer strong generalization, their significant parameter counts and subsequent energy consumption often make them unsuitable for the limited size and power constraints of LEO satellite payloads. Unlike terrestrial data centers, satellites require resource-efficient models that prioritize local traffic patterns over global generalization. However, to provide a competitive analysis, we also utilized a leading foundation model as a benchmark.

\begin{figure*}[t!]
\centerline{\includegraphics[width=1\linewidth]{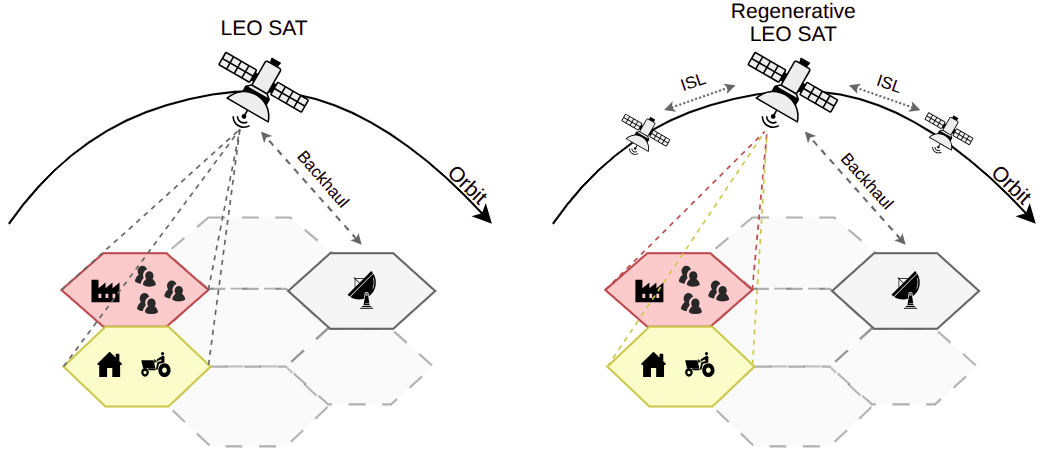}}
\caption{Comparative snapshots of LEO satellite network configurations: On the left, a conventional bent-pipe architecture employing static beam patterns regardless of fluctuating traffic demands. On the right, regenerative system utilizing Inter-Satellite Link (ISL) connectivity and BH to dynamically adapt to real-time user requirements.}
\label{fig:setting}
\end{figure*}

Beyond the choice of forecasting model, 
the user traffic model and its simulation is another 
challenge to be addressed. In the survey work \cite{jiang2023network}, 
the authors shared some tools to generate satellite network traffic. While platforms such as Satellite Network Simulator 3 (sns-3) \cite{sns3} and Open Source Satellite Simulator (OS$^3$) \cite{os3} offer diverse tools for generating satellite traffic, they still necessitate user-defined assumptions regarding the underlying probability distributions. Highlighting this we make the following observations; from the satellite's perspective, user demand is highly aggregated within a single beam. Furthermore, the underlying traffic is fundamentally Internet Protocol (IP) based. Last but not the least, user demand fluctuates over millisecond intervals, producing bursty patterns. This millisecond granularity is particularly important; since the DT can be in millisecond windows, modeling at this resolution enables the system to allocate dwells more precisely to meet rapidly changing user demands. 

Consequently, we assume the underlying traffic model is self-similar. This assumption is supported by three empirical studies across various networking domains. Leland et al. \cite{leland2002self}, established the self-similar nature of Ethernet packets by analyzing Bellcore Labs datasets that spanned up to 40 hours with second granularity. In \cite{paxson}, the authors analyzed WAN packets in a university campus where the traces had up to 3.7M TCP connections which exhibit asymptotic self-similarity. Furthermore,  in \cite{crovella} the authors confirmed these characteristics in high-resolution WWW request traces. Despite this technical consensus, the LEO satellite community currently lacks publicly available traffic datasets to empirically verify these patterns. To bridge this gap and provide a rigorous benchmark for our forecasting framework, we leverage a publicly available WiFi dataset that was collected from a large university campus with a high-density user population. By aggregating traces from hundreds of concurrent users, this dataset serves as a proxy for the bursty, self-similar nature of aggregated LEO beam traffic, allowing for model validation against real-world user traffic pattern. 

Building upon our previous work \cite{demirci}, we introduced several significant extensions.
First, we verified daily user demand fluctuations using WiFi data, which further revealed 
a multi-day seasonality pattern. To strengthen our evaluation, we integrated an 
optimal forecasting benchmark that provides a more robust comparison than the FARIMA 
model used previously. Furthermore, to address the substantial model size of 
Transformers, we introduced a more computationally efficient, smaller-scale neural network 
model. Additionally, we also introduced a foundation transformer model to leverage the strong 
zero-shot performances seen in other domains \cite{aksugift}. We also embedded the
 forecasters in a BH simulator, so that the models are judged by the loss ratio and
 the buffer backlog they induce rather than by predictive accuracy alone. Finally, we
 adopted the Mean Absolute Scaled Error (MASE) instead of standard MSE, as its
 normalization by the persistence forecast makes the results comparable across
 datasets with different scales and burstiness levels, while also indicating whether
 a model provides a genuine gain over simply repeating the last observation.
More particularly, we make the following contributions in this work:
\begin{enumerate}[label=C\arabic*]
    \item We provide a comprehensive comparison of seven forecasters spanning a wide
    spectrum of available solutions, from theoretical optimal benchmarks to SOTA
    foundation transformer models, evaluated on synthetic self-similar traces at three
    levels of long range dependence and on an empirical Wi-Fi trace.
    \item We validate the self-similar traffic model against an empirical high-density
    Wi-Fi trace, and show that the model ranking nearly reverses between the raw and the
    deseasonalized series. The classical forecasters therefore do not fail on empirical
    traffic as such, they require the periodic component to be removed first, which
    reframes the gap as a preprocessing problem rather than a lack of model capacity.
    \item We embed the forecasters in a BH simulator and show, across five scenarios,
    that the accuracy differences barely propagate to the loss ratio and the buffer
    backlog. System utilization and the planning period dominate both metrics. The effect of the
    planning period is moreover non-monotone: below a certain threshold, the rigid
    granularity of the slot allocation, rather than the forecast itself, becomes the
    binding constraint. We show that pursuing pure predictive accuracy is not
    necessarily the most effective route to improving system performance.
\end{enumerate}.
The remainder of this paper is organized as follows: Section II details the considered traffic models and the proposed forecasting models. Section III presents the forecasting accuracy results and the beam hopping simulation study. Finally, Section IV concludes the paper and proposes potential directions for future research.

\section{System Model and Theoretical Framework}

\subsection{The Self-similar Traffic Model}
\label{chap:ssm}
In this work, we consider a LEO satellite constellation with a focus on the forwardlink (downlink). We assume the satellites have regenerative payload and provide broadband services to Earth-fixed cells. Due to the varying number of active users, their heterogeneous demands, and the aggregated nature of traffic at the beam level, we assume that the traffic has self-similar characteristics. More formally, let $\boldsymbol{A}_t$ define the traffic model, a process given by
\begin{equation}
\boldsymbol{A}_t = mt + \sqrt{\alpha m}\boldsymbol{Z}_t, \text{ } t \in (-\infty, \infty)
\label{eq:selfsimilarProcess}
\end{equation}
where $m>0$ is a mean input rate, $\alpha>0$ is a variance coefficient and $H \in [\frac{1}{2},1)$ is the self-similarity parameter of $\boldsymbol{Z}_t$ \cite{norros2}.
 As $\boldsymbol{Z}_t$ is a normalized fractional Brownian motion (fBm), it meets the following properties:
\begin{enumerate}[label=(\roman*)]
    \item $\boldsymbol{Z}$ has stationary increments \label{cond:1}.
    \item $\boldsymbol{Z}_0=0$ and $\mathbb{E}[\boldsymbol{Z}_t]=0$ for $\forall t$ \label{cond:2}.
    \item $\mathbb{E}[\boldsymbol{Z}_t]^2=|t|^{2H}$.\label{cond:3}
    \item $\boldsymbol{Z}$ has continuous paths.
    \item Finite-dimensional distributions of $\boldsymbol{Z}$ are Gaussian distributions. \label{cond:5}
\end{enumerate}

\begin{figure*}[t]
\centerline{\includegraphics[width=1\linewidth]{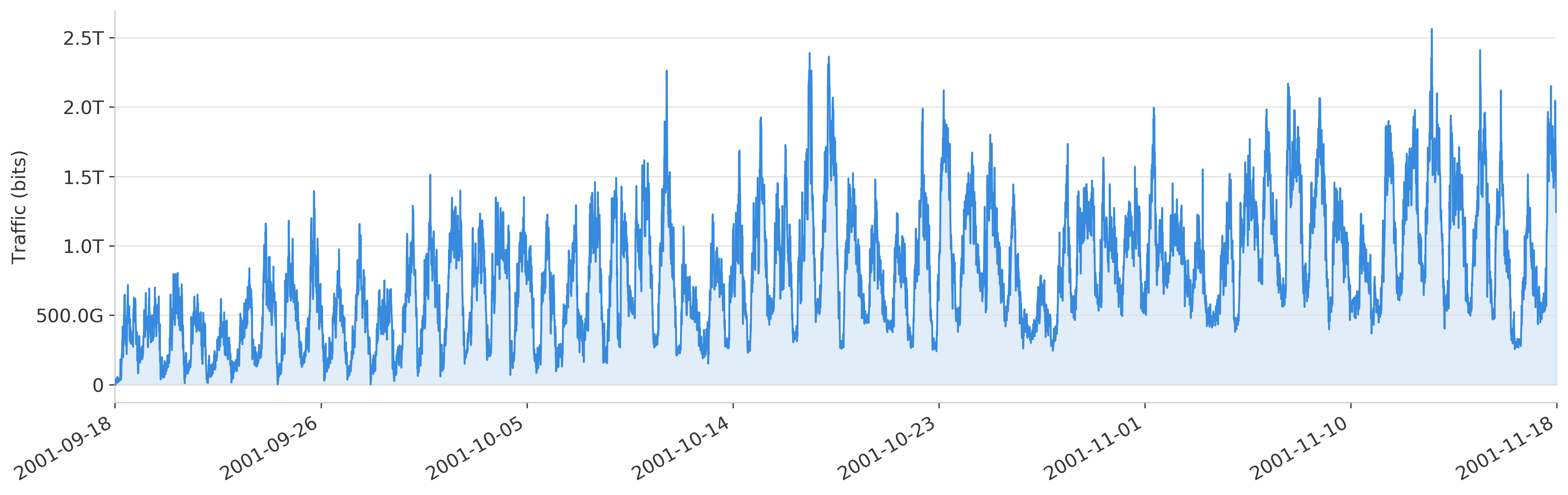}}
\caption{Aggregated downlink traffic demand from the CRAWDAD dataset (Sept. 18 – Nov. 18 2001) \cite{kotz2009crawdad} The SNMP data is aggregated into 10 minute intervals, showing a mean of 359 active users and an average downlink demand of 797 Gbits per interval.}
\label{fig:crawdad}
\end{figure*}

\subsection{Synthetically Generating Self-similar Traffic}
Taqqu et. al \cite{taqquProof} showed that it is possible to generate a self-similar traffic using $M$ Independent and identically distributed (i.i.d.) sources. Following their notation, assume each source generates its own reward sequence $\{ \boldsymbol{W}^{(m)}_t, t\geq0 \}$. The packet generation process is modeled by a binary reward process, $\boldsymbol{W}_t \in \{0, 1\}$, where $\boldsymbol{W}_t=1$ denotes an ON period characterized by packet arrivals and $\boldsymbol{W}_t=0$ denotes an OFF period of inactivity. Given this, consider the superposition of the reward sequences, rescaling time by a factor $T$:
\begin{equation}
    \boldsymbol{W}^{*}_{Tt} = \int_{0}^{Tt} \left( \sum_{m=1}^M \boldsymbol{W}^{(m)}_u \right)du.
\label{eq:superpose}
\end{equation}
When the duration of the ON/OFF periods have a long tailed distribution (such as Pareto), for large $M$ and $T$, the aggregated process $ \{ \boldsymbol{W}^{*}_{Tt}, t\geq0 \}$ behaves statistically like the following as shown in \cite{taqquProof}:
\begin{equation}
    TM\frac{\mu_1}{\mu_1+\mu_2}t+T^H\sqrt{L_{(tM)}}\sigma_{lim}\boldsymbol{Z}_t,
\label{eq:behaveLike}
\end{equation}
where $\mu_i$ equals to $\int_0^\infty xf_i(x)dx$ and 
$f_i(x)$ is the probability density function of ON/OFF periods for
$i=1$ representing ON and $i=2$ representing OFF period. 
Say the probability density function follows a Pareto distribution and both ON and OFF 
periods have the same shape parameter of $\beta$ for $1<\beta<2$.
For both ON and OFF periods, the complementary distribution would be $F_{c}(x) \sim lx^{\beta}L_{(\cdot)}$ where $l$ is a constant and $L_{(\cdot)}$ is a slowly varying function at infinity, $\text{lim}_{x\rightarrow\infty}L_{(tx)}/L_{(x)}=1$ for $t>0$. $\Gamma(\cdot)$ being the Gamma function, $\sigma_{lim}^2$ can be found as \cite{taqquProof}
\begin{equation}
    \sigma_{lim}^2 = \frac{\beta}{2\mu \times \Gamma(4-\beta)}.
\end{equation}

There is a direct relationship between the self-similarity of the traffic model 
and the Hurst parameter as $H=(3-\beta)/2$. Additionally, \eqref{eq:selfsimilarProcess} and 
\eqref{eq:behaveLike} exhibit the same functional structure, as each involves a 
linear scaling of the temporal variable ($t$) and a constant scaling of the 
underlying fBm.

\subsection{Publicly Available Dataset}
To the best of our knowledge, there is no empirical user demand
 datasets for LEO based broadband services. To mitigate this and
rigorously validate our self-similar traffic model, we utilize
a high-density wireless network dataset as a proxy. We argue that
 a Wi-Fi dataset is a reasonable proxy for LEO satellite downlink 
 demand for several reasons. First, in both settings the observed traffic 
 is an aggregation over many users sharing a common access point: 
 for Wi-Fi this aggregation happens at the Acess Point (AP), while for a beam-hopped
LEO satellite it happens at the beam level, so the two traces are structurally
analogous realizations of aggregated access-network demand. 
Second, both systems ultimately deliver a broadband service over IP,
so the resulting traffic inherits similar higher layer statistical 
properties, such as heavy-tailed flow sizes and session durations, 
regardless of the underlying access technology. Third, in both cases
 the last hop is a wireless link with a dynamic, time-varying link budget,
  as opposed to a deterministic wired connection, which is a property
   our self-similar traffic model is meant to capture. We acknowledge
  that the two channels differ in propagation environment and round-trip
time, and that satellite users are more likely to have a Line-of-Sight
(LOS) connection than Wi-Fi users, who are frequently subject to non-LOS
conditions indoors. Nevertheless, both remain wireless links governed
by a variable link budget rather than a fixed-capacity wire, which 
supports using the Wi-Fi trace as a proxy for validating the 
self-similarity of the traffic model.

Specifically, we leverage CRAWDAD dataset from Dathmouth College which has 
records of Simple Network Management Protocol (SNMP) polling 
\cite{kotz2009crawdad}. The data reflects a campus wide network consisting of 
476 Wi-Fi (802.11b) APs serving approximately 5,500 students 
and 1,200 faculty members. For this study, we specifically analyzed the 
``fall01" trace collected during the 2001 Fall term. To analyze downlink traffic, 
we extracted the \textit{awcTpFdbDestOctetsImmed} object and aggregated it per 
AP and MAC address into fixed-length time buckets, handling 32-bit 
counter wraparound and proportionally splitting byte deltas that span 
multiple buckets by time overlap.
\footnote{All source code, along with the processed Wi-Fi dataset, is publicly available at \url{https://github.com/YektaDemirci/leoForecast.git}.} 
While we tracked active user counts, we acknowledge that the dataset 
sanitization process may have affected the precision of these specific values.
To ensure data integrity, we performed a cleaning step to remove ``garbage" entries caused by 
buggy AP firmware as mentioned by the authors. We particularly
considered the data from September 18 to November 18.

Data were aggregated into 10-minute windows to ensure that each interval 
captures at least one polling cycles, given the 5-minute periodicity. While this 
resolution is coarser than our target, it represents the finest granularity 
available among public datasets to the best of our knowledge. For traces spanning 
multiple intervals, we proportionally allocated the octet counts based on 
the duration of the trace within each respective window. 
Figure \ref{fig:crawdad} illustrates the obtained downlink demand data.
 As one can see there is seasonality in the data where the user demand 
 increases during the daytime.

\subsection{Forecasting Models}
We evaluate seven forecasting models which represent a diverse spectrum of solutions, ranging from theoretical statistical predictors to transformer models.

First, we establish a theoretical baseline using the optimal predictor for fractional Brownian motion (fBm) derived by Gripenberg and Norros \cite{norrosForecast}. While specialized, this method provides the mathematically unique optimal forecast for pure fBm traces. To complement this, we employ Fractional ARIMA (FARIMA), as FARIMA processes are known to be asymptotically second-order self-similar \cite{leland2002self}, making them highly suitable for modeling long-range dependent traffic.

Within the of deep learning options, we consider three distinct architectures. We include Informer \cite{zhou2021informer}, which utilizes a \textit{ProbSparse} attention mechanism to mitigate the quadratic time complexity typically associated with vanilla Transformer models. Furthermore, we evaluate the foundation model Chronos \cite{ansari2024chronos}, a time-series foundation model that has demonstrated superior performance across the GIFT-Eval benchmark \cite{aksugift} which is a standardized platform designed for the comparative analysis of foundation TSF models. Finally, we integrate DLinear, a model that challenges the necessity of complex Transformer layers for TSF by proposing a simple, decomposition-based linear architecture that often outperforms more computationally intensive models according to the authors \cite{dlinear}.

\subsubsection{The optimal predictor for continuous fBm}
Despite, being relatively overlooked in the recent literature, Gripenberg and Norros \cite{norrosForecast} derived the closed-form optimal predictor fBm driven processes. Let $\boldsymbol{Z}$ be a fBm process as given in section \ref{chap:ssm}. Then, for a forecasting horizon $\Delta > 0$ and a look-back window $\tau \in (0, \infty]$, the optimal predictor $\boldsymbol{\hat{Z}_{\Delta,\tau}} = \mathbb{E}[\boldsymbol{Z}_\Delta|\boldsymbol{Z}_s, s \in (-\tau,0)]$ is found by Gripenberg and Norros \cite{norrosForecast}:
\begin{equation}
    \boldsymbol{\hat{Z}}_{\Delta,\tau} = \int_{-\tau}^0 g_\tau(\Delta,t) d \boldsymbol{Z}_t \quad \text{for }\tau<\infty, t\in(0,\tau)
\label{eq:forecaster}
\end{equation}
where
\begin{multline}
    g_\tau(\Delta,-t) = \frac{sin(\pi(H-\frac{1}{2}))}{\pi}t^{-H+\frac{1}{2}}(\tau-t)^{(-H+\frac{1}{2})}\\\int_0^\Delta\frac{z^{H-\frac{1}{2}} (z+\tau)^{H-\frac{1}{2}}}{z+t}dz.
\label{eq:smoothFunc}
\end{multline}

It is worth noting \eqref{eq:forecaster} has a stochastic integral. To provide familiarity, the integration is defined as the following for a process of $Y_t=\sum_{k=1}^k \boldsymbol{Z}_t1_{(T_{k-1},T_k)}(t)$ \cite{norrosForecast}:
\[
\int_{-\infty}^\infty \boldsymbol{Y}_t d\boldsymbol{Z}_t  \stackrel{def}{=} \sum_{k=1}^r \boldsymbol{Z}_k (\boldsymbol{Z}_{T_k} - \boldsymbol{Z}_{T_{k-1}}), \quad r>k
\]
Furthermore, it is worth mentioning that the authors found the closed form expression for the relative variance of error $D^2(\boldsymbol{Z}_1-\hat{\boldsymbol{Z}}_{1,\tau})/D^2(\boldsymbol{Z}_1)$ and pointed out that for relatively high $H$ parameter $(>0.9)$ it makes relatively a small difference whether one knows $\boldsymbol{Z}$ on $(-\Delta,0)$ or $(-\infty,0)$ for the prediction of $\boldsymbol{Z}_\Delta$. 

In the remainder of the paper, we refer this forecaster as Norros forecaster.

\subsubsection{The optimal predictor for discrete stationary processes}
\label{chap:linearp}
The predictor in \eqref{eq:forecaster} is optimal with respect to the \emph{continuous} path $\{\boldsymbol{Z}_s, s\in(-\tau,0)\}$. In practice a forecaster observes only the traffic volume accumulated within each measurement interval, so \eqref{eq:smoothFunc} must be collapsed to one weight per interval. We therefore also consider the predictor that is optimal on the discrete grid itself.

Although fBm is self-similar and hence non-stationary, by property \ref{cond:1} of Section \ref{chap:ssm} its increments are stationary. Sampling \eqref{eq:selfsimilarProcess} at unit intervals and normalizing, we obtain the fractional Gaussian noise (fGn) sequence
\begin{equation}
    \boldsymbol{X}_k = \boldsymbol{Z}_k - \boldsymbol{Z}_{k-1}, \quad k \in \mathbb{Z},
\label{eq:fgn}
\end{equation}
which is stationary with zero mean and autocovariance depending only on the lag $k$,
\begin{equation}
    \gamma(k) = \tfrac{1}{2}\left(|k-1|^{2H} - 2|k|^{2H} + |k+1|^{2H}\right).
\label{eq:fgnacov}
\end{equation}

For a sequence of this kind the best linear predictor follows from a classical projection result \cite{brockwell}. Let $S$ and $W_n,\dots,W_1$ be any random variables with finite second moments whose means $\mu_S = \mathbb{E}[S]$, $\mu_i = \mathbb{E}[W_i]$ and covariances are known. Collecting the observations in $\boldsymbol{W} = (W_n,\dots,W_1)'$ with mean vector $\boldsymbol{\mu}_W = (\mu_n,\dots,\mu_1)'$, and writing
\begin{equation}
    \boldsymbol{\gamma} = \mathrm{Cov}(S,\boldsymbol{W}), \qquad
    \boldsymbol{\Sigma} = \mathrm{Cov}(\boldsymbol{W},\boldsymbol{W}),
\label{eq:bdcov}
\end{equation}
the best linear predictor of $S$ in terms of $\{1, W_n,\dots,W_1\}$ \cite{brockwell} is
\begin{equation}
    P(S \mid \boldsymbol{W}) = \mu_S + \boldsymbol{a}'(\boldsymbol{W} - \boldsymbol{\mu}_W),
\label{eq:bdpred}
\end{equation}
where $\boldsymbol{a} = (a_1,\dots,a_n)'$ is any solution of
\begin{equation}
    \boldsymbol{\Sigma}\,\boldsymbol{a} = \boldsymbol{\gamma}.
\label{eq:normaleq}
\end{equation}
Specializing \eqref{eq:bdcov}--\eqref{eq:normaleq} to our notation, let the look-back window be the $T$ most recently observed increments, ordered from newest to oldest,
\begin{equation}
    \boldsymbol{W} = (\boldsymbol{X}_{t},\boldsymbol{X}_{t-1},\dots,\boldsymbol{X}_{t-T+1})',
\label{eq:window}
\end{equation}
and let the target be the demand accumulated over the next $h$ intervals,
\begin{equation}
    S = \boldsymbol{Z}_{t+h} - \boldsymbol{Z}_{t} = \sum_{j=1}^{h}\boldsymbol{X}_{t+j},
\label{eq:target}
\end{equation}
which reduces to $S=\boldsymbol{X}_{t+1}$ for the one-step-ahead case $h=1$. By property \ref{cond:2} both $\mu_S$ and $\boldsymbol{\mu}_W$ vanish, so the intercept term in \eqref{eq:bdpred} drops out. The entries $\Sigma_{ij}$ of $\boldsymbol{\Sigma}$ and $\gamma_i$ of $\boldsymbol{\gamma}$ follow directly from \eqref{eq:fgnacov}:
\begin{equation}
    \Sigma_{ij} = \gamma(i-j), \qquad
    \gamma_i = \sum_{j=1}^{h}\gamma(i + j - 1),
\label{eq:lpsystem}
\end{equation}
for $i,j = 1,\dots,T$. The resulting linear predictor is:
\begin{equation}
    \hat{S}_{h,T} = \boldsymbol{a}'\boldsymbol{W} = \sum_{i=1}^{T} a_i\,\boldsymbol{X}_{t-i+1}.
\label{eq:lppred}
\end{equation}

\eqref{eq:lppred} has a direct consequence for our comparison. Once discretized to the same grid, the Norros forecaster is itself a linear function of the same $T$ increments, and therefore cannot outperform \eqref{eq:lppred}. Any gap between the two is attributable solely to the discretization of \eqref{eq:smoothFunc}, and \eqref{eq:lppred} serves as a strict lower bound on the attainable error. In the remainder of the paper we refer to this predictor as Linear predictor.

\subsubsection{FARIMA}
A FARIMA$(p, d, q)$ process generalizes the standard autoregressive integrated moving average model by allowing the differencing parameter $d$ to take non-integer values. Following the notation of Box and Jenkins \cite{beran2013long}, a FARIMA process $Z_t$ is defined by the equation:
\begin{equation}
    \phi(B)(1-B)^d Z_t = \psi(B)\varepsilon_t, \quad t \geq 1
    \label{eq:farima_main}
\end{equation}
where $\varepsilon_t$ is assumed to be i.i.d. with zero mean and finite variance $\sigma_{\varepsilon}^2$. The term $(1-B)^d$ represents the fractional differencing operator. For $d \in (-\frac{1}{2}, \frac{1}{2})$, this operator is defined via the binomial series expansion:
\begin{equation}
    (1-B)^d = \sum_{j=0}^\infty a_j B^j
    \label{eq:infSeries}
\end{equation}
where the coefficients $a_j$ are determined using the Gamma function $\Gamma(\cdot)$:
\begin{equation}
    a_j = \frac{\Gamma(j-d)}{\Gamma(j+1)\Gamma(-d)} = \frac{\Gamma(d+1)}{\Gamma(j+1)\Gamma(d-j+1)}(-1)^j
\end{equation}
The stationary and invertible nature of the process is governed by the autoregressive polynomial $\phi(k)=1-\sum_{j=1}^p\phi_jk^j$ and the moving-average polynomial $\psi(k)=\sum_{j=0}^q\psi_jk^j$. These polynomials are assumed to have no common roots, with all roots lying outside the unit circle $|k|>1$ \cite{beran2013long}. 

\subsubsection{Informer}
The quadratic time and memory complexity of the self-attention mechanism represents the primary bottleneck when adapting vanilla Transformers for TSF. To mitigate this, the Informer introduces \textit{ProbSparse} self-attention mechanism which identifies and prioritizes the most active queries that contribute significantly to the attention distribution \cite{zhou2021informer}. Given this performance advantage, we considered Informer as the domain specific transformer model in this work.

Let $Q \in \mathbb{R}^{L_Q \times d}$, $K \in \mathbb{R}^{L_K \times d}$ and $V \in \mathbb{R}^{L_V \times d}$ represent the Query, Key and Value matrices respectively, where $d$ represents the model dimension and  $L_Q = L_K = L_V$ represent the input sequence length. The \textit{ProbSparse} attention is defined as:

\begin{equation}
    \mathcal{A}(Q, K, V) = \text{Softmax}\left(\frac{\bar{Q}K^\top}{\sqrt{d_k}}\right)V,
\end{equation}
where $\bar{Q}$ containing the Top-$\bar{L}_Q$ queries of $Q$ under the sparsity  measurement of $M(q,K)$. The number of selected queries is typically set to $u = c \cdot \ln L_Q$, where $c$ is a sampling factor.

The Informer evaluates the importance of a query $q_i$ by measuring the divergence between its attention probability distribution $p(k_j | q_i)$ and a uniform distribution $q(k_j | q_i) = 1/L_K$. Using the Kullback-Leibler (KL) divergence as a metric, a query is considered ``active" if its attention distribution is highly concentrated. The sparsity measurement $M(q_i, K)$ is bounded as follows \cite{zhou2021informer}:
\begin{equation}
    \ln L_K \leq M(q_i, K) \leq \bar{M}(q_i, K) + \ln L_K
\end{equation}
To avoid the $O(L_Q L_K)$ complexity of calculating the full KL-divergence, the model utilizes the \textit{Log-Sum-Exp} approximation for $\bar{M}(q_i, K)$:

\begin{equation}
    \bar{M}(q_i, K) = \max_{j} \left\{ \frac{q_i k_j^\top}{\sqrt{d}} \right\} - \frac{1}{L_K} \sum_{j=1}^{L_K} \frac{q_i k_j^\top}{\sqrt{d}} 
\end{equation}

In practice, to maintain $O(L \log L)$ complexity, the Informer does not compute $\bar{M}$ for all keys. Instead, it randomly samples a subset of keys ($L_K \ln L_Q$) to estimate the measurement. This allows the model to efficiently extract the most informative temporal dependencies while significantly reducing the computational overhead for long-range sequences.

\begin{table}[b!]
\renewcommand{\arraystretch}{1.4} % Better vertical spacing for academic look
\caption{Comparative Summary of Forecasting Architectures}
\label{table_models}
\centering
\small
\begin{tabular}{|l|l|p{3.7cm}|}
\hline
\textbf{Model} & \textbf{Category} & \textbf{Summary}  \\
\hline
Norros & Statistical & Optimal predictor for fBm datasets, utilizing constant weights.  \\
\hline
Linear & Statistical & Best linear predictor of discrete fGn increments from a finite look-back window. \\
\hline
FARIMA & Statistical & Fractional differencing operator $(1-B)^d$ using binomial series and Gamma functions. \\
\hline
Informer & Deep learning & A Transformer model with \textit{ProbSparse} self-attention. \\
\hline
DLinear & Deep learning & Decomposition of (Trend-Seasonal) followed by temporal linear layers. \\
\hline
Chronos & Deep learning & A zero-shot Transformer model underlying T5. \\
\hline
\end{tabular}
\end{table}

\begin{figure*}[t]
\centerline{\includegraphics[width=1\linewidth]{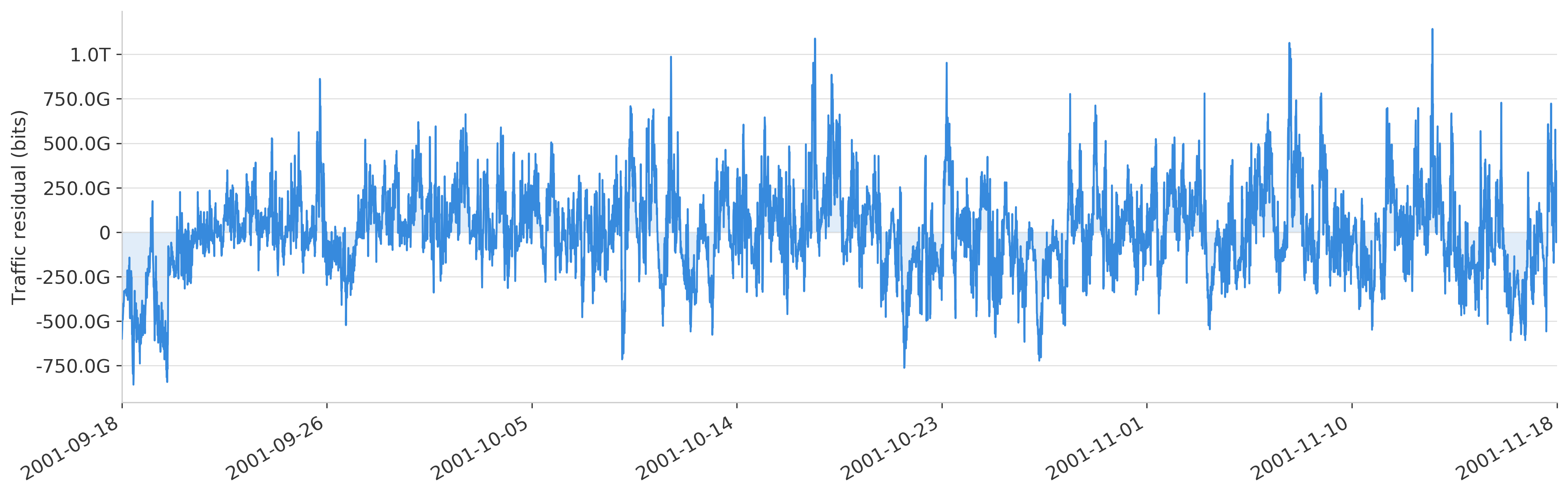}}
\caption{Deseasonalised and detrended downlink demand from the CRAWDAD dataset, obtained from the raw series in Figure \ref{fig:crawdad}.}
\label{fig:deseasonalised}
\end{figure*}

\subsubsection{DLinear}
The authors in \cite{dlinear} introduced DLinear as a surprisingly effective baseline, specifically designed to question the marginal gains provided by relatively heavy transformer architectures for TSF. Unlike the Informer model, which relies on attention to capture dependencies, DLinear utilizes a simple temporal linear layer combined with a decomposition scheme. 
The DLinear model first decomposes the input hidden sequence into two components: a trend-cyclical component ($X_t$) and a seasonal-remainder component ($X_s$). This is achieved through a moving average kernel.

Following the decomposition of the time series, DLinear utilizes a linear layer formulation for each constituent part. This approach regresses the input data directly to the future horizon, leveraging a one-step mapping that bypasses the recursive errors inherent in step-by-step forecasting. As highlighted by the authors, the model maintains a focus on temporal dependencies by sharing weights across different variates rather than explicitly modeling spatial or inter-variate relationships.

Given this simplicity and its documented performance in outperforming Transformer based solutions in certain domains \cite{dlinear}, we include DLinear as one of our benchmark solutions.

\subsubsection{Chronos}
Chronos is a foundation model framework that adapts Transformer based language model architectures from T5 family \cite{raffel2020exploring}, for probabilistic TSF. It is pre-trained on massive and diverse datasets to enable zero-shot generalization across unseen domains. Unlike traditional forecasters that use parametric distributions (like Gaussian), Chronos performs regression via classification. It models a categorical distribution over a fixed vocabulary, which allows it to learn complex, multimodal output distributions without needing specific structural constraints. To map real valued observations into discrete tokens, Chronos uses a two-step tokenization process. Firstly, it employs mean scaling where each individual time series is first normalized using its mean absolute value from the historical context. Secondly, it performs uniform quantization where these scaled values are mapped to one of $B$ discrete bins. The model is then trained on these tokenized sequences using cross-entropy loss.

To achieve strong zero-shot performance, the pre-training process is supplemented by two data augmentation strategies. The first strategy, TSMixup, generates new training samples by taking combinations of randomly sampled scaled time series from different datasets. The second strategy, KernelSynth uses Gaussian processes to generate synthetic time series through a random composition of basis kernels. During inference, Chronos generates forecasts by sampling multiple trajectories from the predicted categorical distribution. These sampled token paths are then de-quantized and unscaled back to the original numerical range. This approach allows the model to handle diverse datasets without requiring task-specific retraining.

\section{Simulations and Numerical Results}

\subsection{Synthetic Traffic Data Generation}
\label{chap:sim}
To model self-similar behavior, we directly generate the aggregate
traffic as the superposition of $M$ independent ON/OFF sources with
Pareto-distributed ON and OFF durations \eqref{eq:superpose}, following the construction of
Taqqu et al. \cite{taqquProof}. Each source alternates between an ON 
period, during which it emits at a fixed rate, and an OFF period of 
silence, with both period lengths drawn from a Pareto distribution with
the same shape parameter $\beta$. The instantaneous aggregate demand is obtained
by summing the emission rate of all $M$ sources at each sampling instant.

We use $M = 500$ sources emitting at $1$~Mb/s while ON and
pareto shape of $\beta \in \{1.04,\, 1.24,\, 1.44\}$, corresponding to $H \in \{0.98,\, 0.88,\,
0.78\}$, with ten independent seeds per value of $\beta$. We used a common scale
$1$ms for both states, so the minimum state durations can be 1ms. Following
Whittle estimation we found the Hurst values of the generated datasets to be
$0.981+-0.015$, $0.875+-0.020$ and $0.782+-0.020 $ for $\beta=1.04$, $\beta=1.24$ and
$\beta=1.44$ respectively. In the remainder of this paper, we refer to these datasets
synthetic datasets.

\subsection{Investigating Self-Similarity in the Wifi Dataset}
As noted in Section \ref{chap:ssm}, the raw CRAWDAD trace exhibits pronounced daily and weekly 
seasonality, which is not accounted for by the self-similar traffic model in 
\eqref{eq:selfsimilarProcess}. We therefore deseasonalise the series with an 
additive decomposition of two nested periods, daily and weekly, inspired by the 
classical seasonal decomposition \cite{brockwell}. For each 10-minute slot we 
subtract a trailing mean over the preceding seven days, which tracks slow drift 
in the diurnal level and shape, followed by a day-of-week profile fitted on the 
training block; both profiles are smoothed across neighbouring slots to suppress 
estimation noise. The weekly profile is estimated on the first 70\% of the 
record and applied unchanged thereafter, while the trailing mean is by 
construction a function of strictly earlier buckets. We followed this split 
because we train our models using the first 70\% of the dataset and didn't want to 
leake information from the test data. Figure \ref{fig:deseasonalised} shows the 
resulting series, in which the daily and weekly cycles of Figure \ref{fig:crawdad} are no longer 
visible.

Using local Whittle estimation \cite{beran2013long}, restricted to the band of frequencies corresponding to periods above 16 hours, 
we obtained a Hurst value of $H=0.84$ for the deseasonalized data. This estimate is representative of a range 
$H \in [0.78, 0.90]$ obtained under nearby, similarly justified bandwidth choices. Following the 
Kwiatkowski-Phillips-Schmidt-Shin (KPSS) test \cite{seabold2010statsmodels} implemented in SciPy, we found a 
$p$-value of $0.1$, which indicates that the null hypothesis of stationarity cannot be rejected. 
This is consistent with the self-similar traffic model in \eqref{eq:selfsimilarProcess}, which is stationary after 
differencing.

We would like to bridge the gap between the 10-minute sampling granularity of the WiFi dataset and the finer 
timescales utilized in subsequent sections. The raw WiFi data could not be resolved at intervals finer than 10 minutes, 
so scale-invariance of the estimated Hurst parameter below this resolution cannot be verified directly from this dataset. 
We instead adopt it as a modeling assumption, consistent with prior evidence that aggregate network traffic exhibits 
long-range dependence over a broad range of timescales \cite{leland2002self}. We note that this assumption 
is a simplification. Building on this, the remainder of this paper models traffic at a finer 
temporal resolution.

\begin{table}[!t]
\centering
\caption{One-step-ahead MASE ($\pm$ 95\% CI over 10 seeds) for the synthetic dataset with $\beta = 1.04$}
\small
\begin{tabular}{|c|l|c|}
\hline
Rank & Forecaster & MASE \\
\hline
1 & \textbf{Linear}     & 0.9237 $\pm$ 0.0093 \\
\hline
2 & Norros      & 0.9248 $\pm$ 0.0086 \\
\hline
3 & FARIMA      & 0.9252 $\pm$ 0.0080 \\
\hline
4 & Chronos*    & 0.9380 $\pm$ 0.0089 \\
\hline
5 & Informer    & 0.9408 $\pm$ 0.0121 \\
\hline
6 & Chronos     & 0.9762 $\pm$ 0.0104 \\
\hline
7 & DLinear     & 1.0880 $\pm$ 0.0253 \\
\hline
\end{tabular}

\label{tab:data1}
\end{table}

\begin{table}[!t]
\centering
\caption{One-step-ahead MASE ($\pm$ 95\% CI over 10 seeds) for the synthetic dataset with $\beta = 1.24$}
\small
\begin{tabular}{|c|l|c|}
\hline
Rank & Forecaster & MASE \\
\hline
1 & \textbf{Linear}     & 0.8840 $\pm$ 0.0081 \\
\hline
2 & FARIMA      & 0.8842 $\pm$ 0.0078 \\
\hline
3 & Norros      & 0.8842 $\pm$ 0.0084 \\
\hline
4 & Chronos*    & 0.8961 $\pm$ 0.0104 \\
\hline
5 & Informer    & 0.8992 $\pm$ 0.0074 \\
\hline
6 & Chronos     & 0.9219 $\pm$ 0.0096 \\
\hline
7 & DLinear     & 0.9662 $\pm$ 0.0158 \\
\hline
\end{tabular}

\label{tab:data2}
\end{table}

\begin{table}[!t]
\centering
\caption{One-step-ahead MASE ($\pm$ 95\% CI over 10 seeds) for the synthetic dataset with $\beta = 1.44$}
\small
\begin{tabular}{|c|l|c|}
\hline
Rank & Forecaster & MASE \\
\hline
1 & \textbf{Linear}     & 0.8393 $\pm$ 0.0030 \\
\hline
2 & \textbf{FARIMA}      & 0.8393 $\pm$ 0.0029 \\
\hline
3 & Norros      & 0.8398 $\pm$ 0.0032 \\
\hline
4 & Informer    & 0.8463 $\pm$ 0.0040 \\
\hline
5 & Chronos*    & 0.8492 $\pm$ 0.0030 \\
\hline
6 & DLinear     & 0.8670 $\pm$ 0.0058 \\
\hline
7 & Chronos     & 0.8711 $\pm$ 0.0054 \\
\hline
\end{tabular}

\label{tab:data3}
\end{table}

\subsection{Implementation and Model Sizes}

We implemented Norros in Python using the integrate function of the SciPy library
\cite{scipy}, and Linear likewise in Python with SciPy, while the FARIMA model 
was executed in R via the arfima package \cite{veenstra2013persistence}. We trained our deep 
learning models using the authors' open source implementations for Informer \cite{zhou2021informer} 
and DLinear \cite{dlinear} respectively. We selected the chronos-bolt-tiny 
variant as our foundation model because it is the smallest available version, making 
it more acceptable for the limited computational resources of LEO satellites. We 
evaluated this model in two ways: first, in its original zero-shot state (Chronos), 
and second by fine-tuning it on our specific traffic datasets (Chronos*).

Training, finetuning and forecasting of the deep learning models were performed 
on an NVIDIA RTX 500 Ada GPU (4 GB memory). The model complexity varies significantly. 
Informer and chronos-bolt-tiny utilize approximately 11.3M and 8.6M trainable parameters, 
respectively. Conversely, DLinear maintains an exceptionally small footprint compared to 
the transformer models, ranging from 150 to 3,000 parameters.

\subsection{Performance Metrics}
To assess accuracy, we utilized the Mean Absolute Scaled Error (MASE), 
defined as: $$MASE = \frac{\frac{1}{n}\sum_{i=1}^{n} |y_i - \hat{y}i|}{\frac{1}{n-1}\sum{i=2}^{n} |y_i - y_{i-1}|},$$
where $y_i$ represents the ground truth and $\hat{y}_i$ 
is the forecasted value. The denominator is the mean absolute 
one-step change of the ground truth over the scored window,
the error of the persistence (naive) forecast. This metric provides 
a scale independent assessment in which $MASE = 1$ corresponds to persistence 
and values closer to 0 indicate higher precision. For all models, the forecasting
horizon is generated based on the 48 most recent demand samples.

\begin{table}[!t]
\centering
\caption{One-step-ahead MASE across different forecasters for the parsed WiFi dataset, on the raw and the deseasonalized series}
\small
\begin{tabular}{|c|l|c||c|l|c|}
\hline
\multicolumn{3}{|c||}{Raw} & \multicolumn{3}{c|}{Deseasonalized} \\
\hline
Rank & Forecaster & MASE & Rank & Forecaster & MASE \\
\hline
1 & \textbf{DLinear}   & 0.9864 & 1 & \textbf{FARIMA}    & 0.9665 \\
\hline
2 & Chronos*  & 0.9873 & 2 & DLinear   & 0.9756 \\
\hline
3 & Informer  & 1.0229 & 3 & Chronos*  & 0.9903 \\
\hline
4 & FARIMA    & 1.0549 & 4 & Norros    & 0.9922 \\
\hline
5 & Chronos   & 1.0659 & 5 & Informer  & 0.9966 \\
\hline
6 & Norros    & 1.1278 & 6 & Linear   & 1.0011 \\
\hline
7 & Linear   & 1.1690 & 7 & Chronos   & 1.0528 \\
\hline
\end{tabular}

\label{tab:wifi}
\end{table}
 
\subsection{Forecasting Accuracy Results}
The forecasting performance of the evaluated models is summarized in
Tables \ref{tab:data1} through \ref{tab:wifi}, reported as one-step-ahead MASE
so that a value of $1$ corresponds exactly to the persistence forecast and
values below $1$ indicate a genuine gain over it. Tables \ref{tab:data1} to
\ref{tab:data3} detail the results for the synthetic datasets, where we vary
the Pareto shape parameter $\beta$ to observe the impact of different
self-similarity intensities; each entry is the mean over $10$ seeds
together with the half-width of the $95\%$ confidence interval across seeds.
Table \ref{tab:wifi} then presents
the performance results using empirical WiFi traffic data, evaluated both on the
raw series and on the deseasonalized series of Figure \ref{fig:deseasonalised}. 
Our analysis focuses on two main
goals: (i) understanding how self-similarity affects forecasting accuracy and
(ii) assessing how well the models generalize from synthetic to real-world data.

In our synthetic datasets, as $\beta$ increases from 1.04 to 1.44, the intensity of
Long Range Dependency (LRD) decreases. Correspondingly, the best MASE value, attained by the Linear
forecaster of Section \ref{chap:linearp}.
The MASE values comparing the datasets with different $\beta$ must be interpreted
cautiously, as the numerical values themselves do not imply that strongly
self-similar traffic is inherently more difficult to forecast. Because MASE is
normalized by the persistence forecast, it reports the relative gain over
repeating the last observation rather than the absolute forecast error. In
datasets with strong LRD (higher Hurst parameter $H$), the persistence forecast
itself is highly accurate due to strong positive autocorrelation
\eqref{eq:fgnacov}.

As shown in Tables \ref{tab:data1}, \ref{tab:data2}, and \ref{tab:data3}, the
three self-similarity aware forecasters, Linear, Norros and FARIMA, occupy the
top three ranks at every value of $\beta$, and their confidence intervals overlap
throughout. Linear is nominally first in all three settings, which is expected:
it is the predictor that is optimal on the discrete sampling grid
(Section \ref{chap:linearp}), whereas the Norros weights of \eqref{eq:smoothFunc}
are optimal for the continuous path and lose a small amount of accuracy when
collapsed to one weight per measurement interval. This is the cost of
discretization. At $\beta=1.44$ the Linear and FARIMA forecasters in fact reach the
same MASE of $0.8393$. This convergence is not
coincidental. All three are, by different routes, estimates of the same optimal
linear predictor for fGn. The practical implication is 
that on purely self-similar traffic these three forecasters are interchangeable,
and the choice between them should be governed by implementation cost rather 
than accuracy.

The learned models trail this group consistently. Fine-tuning is what makes the
foundation model competitive at all, moving Chronos* ahead of the zero-shot
Chronos, yet Chronos* never reaches the top three
and Informer ranks fourth or fifth throughout. DLinear is the weakest model on
synthetic traces.

When evaluated against the WiFi dataset (Table \ref{tab:wifi}), the
picture changes substantially. On the raw series DLinear has the best accuracy, 
and the models that dominated the synthetic experiments are now the worst 
performers Linear and Norros. The ordering is close to a reversal of the synthetic
ranking, and the cause is the seasonality visible in Figure \ref{fig:crawdad}:
the time-of-day and day-of-week profile violates the stationary increment
assumption that  forecasters are derived from.

Investigating the deseasonalized data confirms the claim. On the deseasonalized series every
model except Chronos* improves, and the gains are concentrated precisely in the
self-similarity aware forecasters. FARIMA becomes the best model overall at 
$0.9665$. The classical forecasters
therefore do not fail on empirical traffic as such; they require the periodic
component to be removed first, whereas the deep models absorb part of it
internally and consequently benefit far less from the same preprocessing.

These accuracy differences come with different computational costs. Informer is
computationaly expensive compared to Linear and Norros forecasters: its model size 
is large and its run time complexity is $O(L \log L)$. 
Chronos* adds further overhead, since it also requires
fine-tuning on representative traffic before it becomes competitive. Considering Norros forecaster, 
the weight function in
\eqref{eq:smoothFunc} depends only on the Hurst parameter $H$, the look-back
window $T$, and the forecasting horizon $\Delta$.

These results point to a practical trade-off driven primarily by data preparation rather than model size. 
Under specific operational conditions involving targeted signal preprocessing, specialized lightweight models can 
achieve performance comparable to larger, more complex architectures while significantly reducing computational overhead.

\begin{table*}[!b]
\centering
\caption{Configuration of the five beam hopping scenarios. Every scenario uses three cells sharing one beam.}
\small
\begin{tabular}{|c|l|c|c|c|c|c|}
\hline
\makecell{Scenario} & \makecell{Varied\\parameter} &
\makecell{Beam capacity\\$C$ [Mb/s]} & $\rho = 750/C$ &
\makecell{Forecast period\\ \& Granularity} &
\makecell{Sources per cell\\(cell 1 / 2 / 3)} &
\makecell{Pareto shape $\beta$\\(cell 1 / 2 / 3)} \\
\hline
1 & \makecell[l]{System\\utilization $\rho$} &
\makecell{760\\790\\833\\938} & \makecell{0.99\\0.95\\0.90\\0.80} &
1.5 & 500 / 500 / 500 & 1.04 / 1.04 / 1.04 \\
\hline
2 & \makecell[l]{Forecasting\\granularity} &
790 & 0.95 & \makecell{0.015\\0.15\\1.5\\15\\60} &
500 / 500 / 500 & 1.04 / 1.04 / 1.04 \\
\hline
3 & \makecell[l]{Asymmetric demand\\self-similarity} &
790 & 0.95 & 1.5 & 500 / 500 / 500 & 1.04 / 1.24 / 1.44 \\
\hline
4 & \makecell[l]{Asymmetric\\number of users} &
790 & 0.95 & 1.5 & 950 / 450 / 150 & 1.04 / 1.04 / 1.04 \\
\hline
5 & \makecell[l]{Asymmetric users and\\demand self-similarity} &
790 & 0.95 & 1.5 & 950 / 450 / 150 & 1.04 / 1.24 / 1.44 \\
\hline
\end{tabular}

\label{tab:scenarios}
\end{table*}

\subsection{Beam Hopping Simulation Scenarios}

After analysing the forecasting accuracy of the models in isolation, 
we investigate what that accuracy is worth to a BH scheduler. We implemented in 
Python a BH simulator that abstracts away the radio and mobility layers and 
retains only the elements the forecast acts on: demand generation, the periodic 
illumination plan, and the per-cell buffers it feeds. Three cells share a single 
beam, at 1 ms slot granularity. Each cell's demand is an independent superposition 
of Pareto ON/OFF sources, which yields self-similar traffic, and is accumulated into 
a per-cell buffer of 1 MiB; arrivals that do not fit are dropped, and the 
resulting loss ratio is one of our metrics. The total generated demand is created by 1500 sources across every scenario.

Under fixed-period scheduling, a plan is generated from a one-step-ahead 
forecast of each cell's demand for the upcoming period. Each cell's weight is 
its forecast demand, clipped at zero, plus its current buffer occupancy. The slots are then apportioned to the 
cells in proportion to these weights
using largest-remainder apportionment: the exact real-valued shares are
floored, and the leftover slots go to the cells with the largest fractional
parts. Finally, 
the resulting per-cell slot counts are interleaved into the beam schedule 
round-robin rather than as contiguous blocks, so the cells are visited more 
evenly, reducing the peak buffer occupancy down. In each slot, the illuminated 
cell drains the served amount of bits from its buffer, capped by what it
currently holds.

Two forecasters are compared on the identical traffic realisation, 
so that any difference between them is attributable solely to the forecaster: 
Linear and DLinear. These two models were selected because they represent 
opposite ends of the performance spectrum on the synthetic datasets, 
with one achieving the highest accuracy and the other the lowest.
Each scenario is run for ten independent replicates, 
with each replicate utilizing a testing window spanning two hours of traffic. 
The models are refitted per replicate using an independent training realization 
of 4,200 samples. To ensure a fair comparison, all scenarios share a common seed 
base, allowing policies to be evaluated pairwise on identical traffic traces. 

We investigate three metrics: i. the loss ratio, ii. the mean backlog, and 
iii. the maximum mean per-cell backlog.

i. The loss ratio is the total number of bits dropped across all three cells 
divided by the total bits offered over the whole test window. 
A drop is recorded at arrival time: at each millisecond, a cell's occupancy plus 
that millisecond's arrivals is clipped to the 1 MiB buffer capacity and 
any excess is discarded and counted. 

ii. The mean backlog Mean backlog is the time average over every 1 ms slot of 
the test window of the total bits queued across the three cell buffers, in Mbit. 
It measures how loaded the buffers run, so it separates planners even when 
nothing overflows and the loss ratio is zero.

iii. The maximum mean per-cell backlog is the mean backlog of the worst-served
cell: for each cell, we compute the time-averaged occupancy of its buffer
exactly, from every-slot accumulators, and then take the maximum over the
three cells.

We evaluate five distinct scenarios to investigate the impact of forecasting 
alongside various network design choices. While every experiment instantiates 
exactly three cells, we systematically vary the beam's service rate 
(system utilization), the forecasting period, the number 
of sources per cell, and the burstiness of the underlying traffic 
(Hurst parameter) as shown in Table \ref{tab:scenarios}.

\subsubsection{Scenario 1: Varying system utilization}

\begin{table}[!t]
\centering
\caption{Scenario 1 varying system utilization from 0.98 to 0.80.}
\small
{\setlength{\tabcolsep}{2pt}\renewcommand{\arraystretch}{1.15}
\begin{tabular}{|c|l|c|c|c|}
\hline
\makecell{$\rho$\\($C$ [Mb/s])} & Policy &
\makecell{Loss ratio\\{[ppm]}} & \makecell{Backlog mean\\{[Mb]}} & \makecell{Max mean\\per-cell [Mb]} \\
\hline
\multirow{2}{*}{\makecell{0.98\\(760)}} &
Linear  & $2499 \pm 1000$ & $6.56 \pm 2.00$ & $2.22 \pm 0.70$ \\
& DLinear & $2524 \pm 1000$ & $6.53 \pm 2.00$ & $2.36 \pm 0.80$ \\
\hline
\multirow{2}{*}{\makecell{0.95\\(790)}} &
Linear  & $2.68 \pm 2.0$ & $0.90 \pm 0.06$ & $0.32 \pm 0.02$ \\
& DLinear & $4.30 \pm 4.0$ & $0.92 \pm 0.07$ & $0.33 \pm 0.03$ \\
\hline
\multirow{2}{*}{\makecell{0.90\\(833)}} &
Linear  & $0$ & $0.78 \pm 0.01$ & $0.27 \pm {<}0.01$ \\
& DLinear & $0$ & $0.78 \pm 0.01$ & $0.27 \pm {<}0.01$ \\
\hline
\multirow{2}{*}{\makecell{0.80\\(938)}} &
Linear  & $0$ & $0.76 \pm 0.01$ & $0.26 \pm {<}0.01$ \\
& DLinear & $0$ & $0.76 \pm 0.01$ & $0.26 \pm {<}0.01$ \\
\hline
\end{tabular}}

\label{tab:exp0}
\end{table}

\begin{table}[!t]
\centering
\caption{Scenario 2, varying forecasting periods \& granularity from 15ms to 60s.}
\small
{\setlength{\tabcolsep}{2pt}\renewcommand{\arraystretch}{1.15}
\begin{tabular}{|c|l|c|c|c|}
\hline
\makecell{Period\\{[s]}} & Policy &
\makecell{Loss ratio\\{[ppm]}} & \makecell{Backlog mean\\{[Mb]}} & \makecell{Max mean\\per-cell [Mb]} \\
\hline
\multirow{2}{*}{0.015} &
Linear  & $0.65 \pm 0.6$ & $1.02 \pm 0.06$ & $0.37 \pm 0.03$ \\
& DLinear & $0.69 \pm 0.6$ & $1.02 \pm 0.06$ & $0.37 \pm 0.03$ \\
\hline
\multirow{2}{*}{0.15} &
Linear  & $0.70 \pm 0.6$ & $0.91 \pm 0.05$ & $0.32 \pm 0.02$ \\
& DLinear & $0.77 \pm 0.7$ & $0.91 \pm 0.06$ & $0.33 \pm 0.03$ \\
\hline
\multirow{2}{*}{1.5} &
Linear  & $3.82 \pm 3$ & $0.99 \pm 0.09$ & $0.35 \pm 0.03$ \\
& DLinear & $6.51 \pm 5$ & $1.01 \pm 0.10$ & $0.38 \pm 0.05$ \\
\hline
\multirow{2}{*}{15} &
Linear  & $51.7 \pm 30$ & $1.17 \pm 0.20$ & $0.42 \pm 0.07$ \\
& DLinear & $60.4 \pm 40$ & $1.19 \pm 0.20$ & $0.46 \pm 0.09$ \\
\hline
\multirow{2}{*}{60} &
Linear  & $89.7 \pm 50$ & $1.26 \pm 0.20$ & $0.46 \pm 0.08$ \\
& DLinear & $101.4 \pm 60$ & $1.28 \pm 0.20$ & $0.51 \pm 0.10$ \\
\hline
\end{tabular}}

\label{tab:exp1}
\end{table}

Table \ref{tab:exp0} reports the three metrics as the beam capacity is swept
and the system utilization falls from
$\rho = 0.983$ to $\rho = 0.797$. All three metrics are dominated by the
utilization rather than by the choice of forecaster. Table \ref{tab:exp0} clearly
shows that under low utilization, the forecasting accuracy is irrelevant given the considered 
metrics. The reason is that at low utilization, the beam has enough spare capacity to fully process 
mis-forecasted cells within a single period, emptying the queue before forecast 
errors can accumulate into a backlog.
\subsubsection{Scenario 2: Varying forecasting granularity}

Table~\ref{tab:exp1} sweeps the planning period from 15 ms to 60 s at a fixed 
utilization. The loss ratio is essentially flat below 150 ms, and then grows by two orders of magnitude. 
The mean backlog is not monotone: it is lowest at 150 ms, slightly higher at 15 ms, and rises to $1.26$ Mb at 60 s; 
the maximum mean per-cell backlog 
follows the same shape, with its minimum at 150 ms. The two 
forecasters are indistinguishable for periods of 150 ms and below, while at 
longer periods Linear attains the lower loss ratio, although the replicate 
standard deviations of the two overlap throughout.

\subsubsection{Scenario 3: Different self similarity characteristics}

\begin{table}[!t]
\centering
\caption{Scenario 3, Varying self-similarity characteristics where all cells either has a Hurst parameter of 0.98, 0.88, or 0.78.}
\small
{\setlength{\tabcolsep}{2pt}\renewcommand{\arraystretch}{1.15}
\begin{tabular}{|c|l|c|c|c|}
\hline
\makecell{$\beta$\\($H$)} & Policy &
\makecell{Loss ratio\\{[ppm]}} & \makecell{Backlog mean\\{[Mb]}} & \makecell{Max mean\\per-cell [Mb]} \\
\hline
\multirow{2}{*}{\makecell{1.04\\(0.98)}} &
Linear  & $2.68 \pm 2.0$ & $0.90 \pm 0.06$ & $0.32 \pm 0.02$ \\
& DLinear & $4.30 \pm 4.0$ & $0.92 \pm 0.07$ & $0.33 \pm 0.03$ \\
\hline
\multirow{2}{*}{\makecell{1.24\\(0.88)}} &
Linear  & $0.056 \pm 0.06$ & $0.86 \pm 0.01$ & $0.29 \pm {<}0.01$ \\
& DLinear & $0.059 \pm 0.06$ & $0.86 \pm 0.01$ & $0.29 \pm {<}0.01$ \\
\hline
\multirow{2}{*}{\makecell{1.44\\(0.78)}} &
Linear  & $0$ & $0.77 \pm {<}0.01$ & $0.26 \pm {<}0.01$ \\
& DLinear & $0$ & $0.77 \pm {<}0.01$ & $0.26 \pm {<}0.01$ \\
\hline
\end{tabular}}

\label{tab:exp2}
\end{table}

\begin{table}[!t]
\centering
\caption{Scenario 4 where cells have 950, 450, 150 users respectively.}
\small
{\setlength{\tabcolsep}{4pt}\renewcommand{\arraystretch}{1.15}
\begin{tabular}{|l|c|c|c|}
\hline
Policy &
\makecell{Loss ratio\\{[ppm]}} & \makecell{Backlog mean\\{[Mb]}} & \makecell{Max mean\\per-cell [Mb]} \\
\hline
Linear  & $0.067 \pm 0.07$ & $1.00 \pm 0.02$ & $0.372 \pm 0.009$ \\
\hline
DLinear & $0.070 \pm 0.07$ & $1.00 \pm 0.02$ & $0.370 \pm 0.009$ \\
\hline
\end{tabular}}

\label{tab:exp3}
\end{table}

\begin{table}[!t]
\centering
\caption{Scenario 5 where cells have 950, 450, 150 users and Pareto shape $\beta$ of 1.04, 1.24 and 1.44 respectively.}
\small
{\setlength{\tabcolsep}{4pt}\renewcommand{\arraystretch}{1.15}
\begin{tabular}{|l|c|c|c|}
\hline
Policy &
\makecell{Loss ratio\\{[ppm]}} & \makecell{Backlog mean\\{[Mb]}} & \makecell{Max mean\\per-cell [Mb]} \\
\hline
Linear  & $0.042 \pm 0.04$ & $1.02 \pm 0.02$ & $0.371 \pm 0.009$ \\
\hline
DLinear & $0.030 \pm 0.03$ & $1.02 \pm 0.02$ & $0.371 \pm 0.008$ \\
\hline
\end{tabular}}

\label{tab:exp4}
\end{table}

Table \ref{tab:exp2} varies the burstiness of the demand generated by each cell.
The traffic were generated with Pareto shapes of $\beta = 1.04$, $1.24$ and $1.44$
causing different self-similar characteristics.

The gap between the forecasters closes as self similarity decreases. At $H = 0.98$
Linear retains the advantage seen in the previous scenarios, but by $H = 0.78$
the considered metrics are similar. Forecaster choice therefore matters only
where the traffic is bursty enough.

\subsubsection{Scenario 4: Asymmetric number of users with identical self-similarity characteristics.}

Table \ref{tab:exp3} shows that asymmetric number of users does not significantly impact 
the considered metrics: the loss ratio falls to below $0.1$ ppm. The two 
forecasters are indistinguishable considering backlog mean and maximum mean per-cell backlog.

\subsubsection{Scenario 5: Asymmetric number of users with varying self-similarity characteristics.}

Compared to Scenario 4, we also changed the self-similarity characteristics of the
cells alongside their asymmetric number of users. According to \ref{tab:exp4} except the
loss ratio, the performance of both forecasters remains similar.

Across all evaluated scenarios, the difference in accuracy between forecasters 
does not translate into a significant change in performance metrics. In fact, 
system utilization and forecasting granularity dictate system performance much more 
than the choice of forecasting model. It is also worth noting that under low utilization, 
the specific forecaster chosen becomes effectively irrelevant. Conversely, 
pushing the system to higher utilization enters regimes where the loss ratio naturally 
increases. Therefore, pursuing pure forecasting accuracy may not be the most effective
strategy for improving overall system performance.

\section{Conclusion}
In this paper, we evaluated statistical and deep learning-based traffic forecasters for proactive capacity 
allocation in LEO satellite networks with BH. We first compared the models in 
isolation by measuring the one-step-ahead MASE, using both synthetic self-similar traces 
and an empirical Wi-Fi trace as a proxy for aggregated beam demand. Finally, we evaluated 
them in context using a BH simulator where the forecast drives the illumination plan. Our 
results show that the forecasting accuracy does not translate into system performance metrics.

On purely self-similar traffic, the three self-similarity aware forecasters,
Linear, Norros and FARIMA, occupied the top three ranks at every value of the
Pareto shape parameter. On the raw Wi-Fi trace the ranking was close to a reversal: DLinear was the most
accurate one alongside the deep learning models. Removing the seasonality, FARIMA became 
the best model overall and both statistical and deep learning based models performed
comparably.

Next, using a BH simulator, we evaluated how the forecasters impact broader 
system metrics. Across all five scenarios, the substantial accuracy gap between 
the best and worst synthetic forecasters (Linear and DLinear) did not translate
 into a comparable difference in loss ratio, mean backlog, or worst-cell mean 
 backlog. Rather, system utilization and forecasting granularity proved to 
 dictate overall performance far more than the choice of forecaster.

Taken together, these findings argue for a design approach that departs from 
the conventional pursuit of raw predictive accuracy. Optimizing the utilization
 margin and carefully tuning the planning period are likely to yield much 
 larger system-level gains than simply deploying a more accurate forecasting 
 model.

In future work, we plan to relax the current abstractions of our simulator by 
incorporating dynamic channel conditions and joint power and bandwidth allocation, 
alongside expanding our analysis to a full constellation level. Additionally, 
evaluating power consumption, memory usage, and overall scalability will be a 
key priority.

\section*{Acknowledgment}
This work was supported in part by MDA Space; in part by the Consortium de Recherche et d’innovation en Aérospatiale au Québec (CRIAQ); and in part by the Natural Sciences and Engineering Research Council of Canada (NSERC).
\bibliographystyle{IEEEtran} 
\bibliography{main}

% ---------------------------------------------------------------------------
% Author biographies. Each block currently shows an empty placeholder frame of
% the correct size; uncomment the \includegraphics line and delete the
% \framebox line once the photo file is available. For a biography with no
% photo at all, use \begin{IEEEbiographynophoto}{Name} with no optional
% argument.
% ---------------------------------------------------------------------------

\begin{IEEEbiography}[{%
% Replace the framebox with the photo (1in x 1.25in, >=300 dpi):
\includegraphics[width=1in,height=1.25in,clip,keepaspectratio]{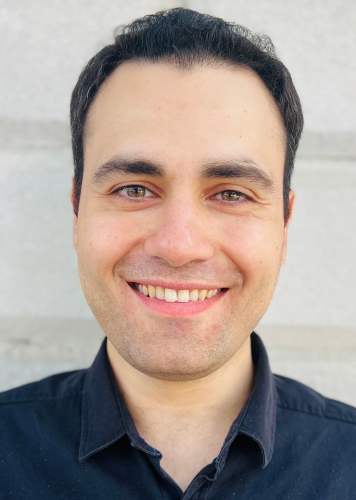}
% \framebox[1in]{\rule{0pt}{1.25in}}%
}]{Yekta Demirci}
(Student Member, IEEE) received the B.Sc. degree in electrical and electronics
engineering from Middle East Technical University (METU), Ankara, T\"urkiye, in 2019
(with honors), and the M.A.Sc. degree in electrical and computer engineering from the
University of Waterloo, Waterloo, ON, Canada, in 2022. From 2022 to 2023, he was
a Full Stack Software Engineer with a Toronto-based startup. He is currently
pursuing the Ph.D. degree with the Poly-Grames Research Center, Polytechnique
Montr\'eal, QC, Canada. His research interests include beam management, traffic
demand forecasting, and RF channel modeling for LEO satellite systems.
\end{IEEEbiography}

\begin{IEEEbiography}[{%
% Replace the framebox with the photo (1in x 1.25in, >=300 dpi):
\includegraphics[width=1in,height=1.25in,clip,keepaspectratio]{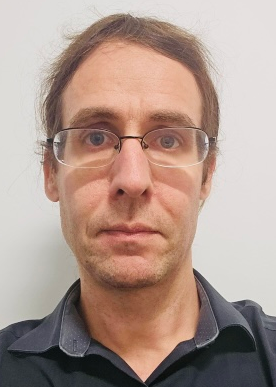}
% \framebox[1in]{\rule{0pt}{1.25in}}%
}]{Guillaume Mantelet}
received the equivalent to M.Sc. in
telecommunication engineering from ENST Bretagne
(now IMT Atlantique) in 2007, and Ph.D in electrical
engineering from the ETS, Canada, in 2012.
His research interest are next-generation network
communications. Since 2023 he has been a Senior
Software Developer at Satellite Systems, MDA,
where he contributes to the development of
high‑performance software for space‑based
communications.
\end{IEEEbiography}

\begin{IEEEbiography}[{%
% Replace the framebox with the photo (1in x 1.25in, >=300 dpi):
\includegraphics[width=1in,height=1.25in,clip,keepaspectratio]{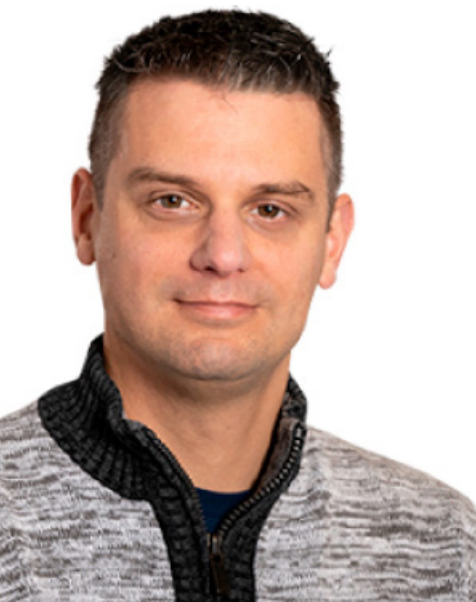}
% \framebox[1in]{\rule{0pt}{1.25in}}%
}]{St\'ephane Martel}
received the bachelor’s
degree in electrical engineering from McGill
University in 2001. He was with the broadcast
television industry, until 2021, where he
occupied multiple roles in software development
and management. He specializes in network
communications and embedded software
development. He joined Satellite Systems, MDA,
in 2021 as a Research and Development Product
Development Manager, where he contributes
to the advancement of space communication
technologies.
\end{IEEEbiography}

\begin{IEEEbiography}[{%
% Replace the framebox with the photo (1in x 1.25in, >=300 dpi):
\includegraphics[width=1in,height=1.25in,clip,keepaspectratio]{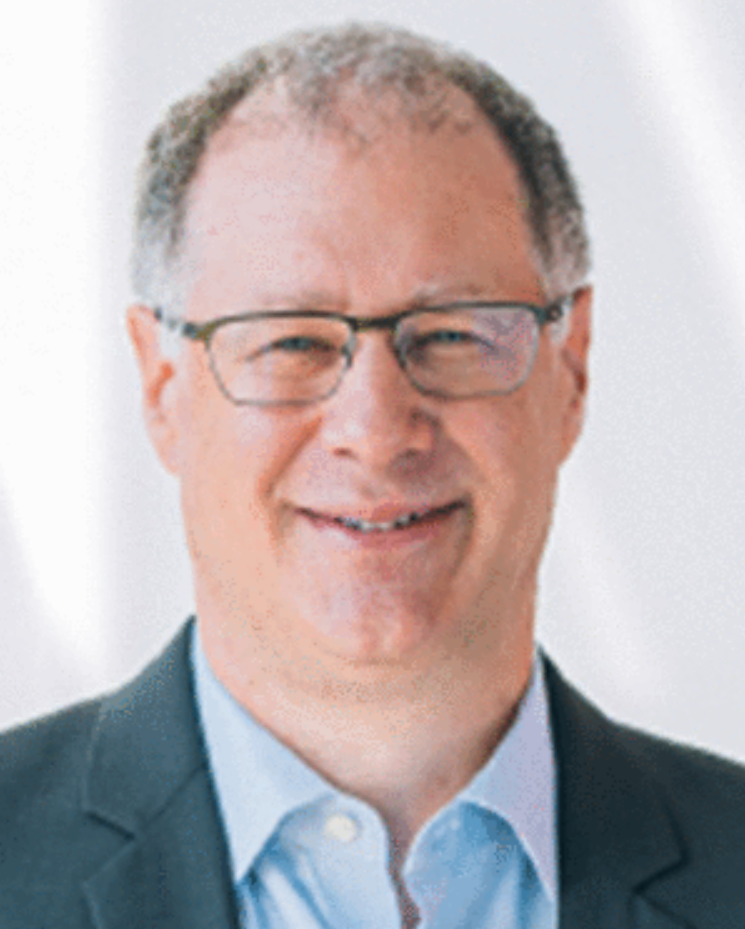}
% \framebox[1in]{\rule{0pt}{1.25in}}%
}]{Jean-Fran\c{c}ois Frigon}
(Senior Member, IEEE) received the B.Eng. degree from the Ecole Polytechnique de Montreal, Montreal, QC, Canada, 
in 1996, the M.A.Sc. degree from the University of British Columbia, Vancouver, BC, Canada, in 1998, and the Ph.D. 
degree from the University of California, Los Angeles (UCLA), Los Angeles, CA, USA, in 2004. 
He joined the Electrical Engineering Department, Ecole Polytechnique de Montreal, in 2004, 
where he is currently a Full Professor and the Department Head. 
He has authored or coauthored more than 115 articles in peer-reviewed journals and conferences and is a co-inventor 
for 13 granted U.S. patents. His research interests include reconfigurable antennas, machine learning algorithms for 
wireless communications, energy harvesting wireless sensor networks, and energy efficient radio resource management in 
wireless systems.
\end{IEEEbiography}

\begin{IEEEbiography}[{%
% Replace the framebox with the photo (1in x 1.25in, >=300 dpi):
\includegraphics[width=1in,height=1.25in,clip,keepaspectratio]{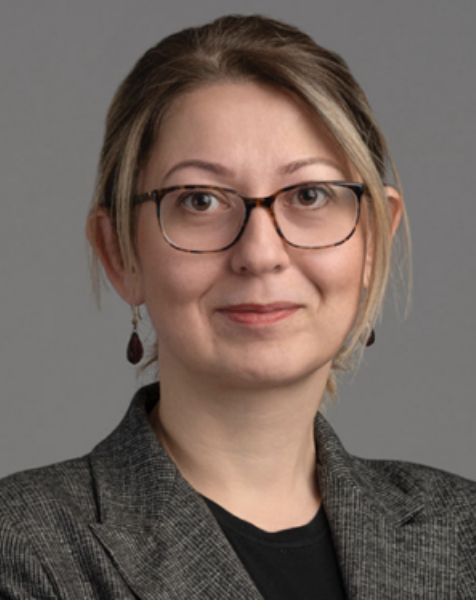}
%\framebox[1in]{\rule{0pt}{1.25in}}%
}]{Gunes Karabulut Kurt}
(Senior Member, IEEE)
received the B.S. degree (Hons.) in electronics and electrical engineering from Bo\u{g}azi\c{c}i University, Istanbul, 
Türkiye, in 2000, and the M.A.Sc. and Ph.D. degrees in electrical engineering from the University of Ottawa, 
Canada, in 2002 and 2006, respectively. She is currently a Canada Research Chair (Tier 1) with the New 
Frontiers in Space Communications and a Professor with Polytechnique Montréal, QC, Canada. She is the 
Director of the Poly-Grames Research Center and the Co-Founder and the Scientific Director of ASTROLITH, 
Transdisciplinary Research Unit of Space Resource and Infrastructure Engineering, Polytechnique Montréal. 
She is also an Adjunct Research Professor with Carleton University, Canada. Prior to her current position, s
he was a Professor with Istanbul Technical University from 2010 to 2021. She is a Marie Curie Fellow and a 
fellow of Canadian Academy of Engineering.
\end{IEEEbiography}

% \begin{IEEEbiographynophoto}{Olfa Ben Yahia}
% received the ... degree in ... . Research interests include ... .

\end{document}